\documentclass[journal, twoside]{IEEEtran}

\usepackage{amsmath}
\usepackage{dsfont}
\usepackage{amssymb,latexsym}
\usepackage{amsfonts}
\usepackage{amscd}
\usepackage{algorithm}
\usepackage{algorithmic}
\usepackage{multirow}
\usepackage{array}
\usepackage{changepage}
\usepackage{times}
\usepackage{float}
\usepackage{graphics,subfigure}
\usepackage{graphbox}
\usepackage[latin1]{inputenc}
\usepackage{graphicx}
\usepackage{color}
\usepackage{relsize,balance}
\usepackage{cite}
\usepackage{stfloats}
\usepackage{epstopdf}
\usepackage{setspace}

\newcommand{\bs}{\boldsymbol}

\newcommand{\mA}{{\bs A}}
\newcommand{\mI}{{\bs I}}
\newcommand{\mK}{{\bs K}}

\newcommand{\mR}{{\bs R}}
\newcommand{\mS}{{\bs S}}
\newcommand{\mP}{{\bs P}}

\newcommand{\mDelta}{{\bs \Delta}}
\newcommand{\mLambda}{{\bs \Lambda}}
\newcommand{\mSigma}{{\bs \Sigma}}
\newcommand{\mPhi}{{\bs \Phi}}
\newcommand{\mPi}{{\bs \Pi}}

\newcommand{\0}{{\bs 0}}
\newcommand{\1}{{\textbf{1}}}

\newcommand{\vs}{{\bs s}}

\newcommand{\vu}{{\bs u}}
\newcommand{\vv}{{\bs v}}
\newcommand{\vw}{{\bs w}}
\newcommand{\vx}{{\bs x}}

\newcommand{\bA}{\bs{\mathcal {A}}}
\newcommand{\bI}{\bs{\mathcal {I}}}

\newcommand{\bR}{\bs{\mathcal {R}}}

\newcommand{\hatvw}{{\widetilde{\bs w}}}
\newcommand{\vpsi}{{\bs \psi}}
\newcommand{\hatvpsi}{{\widetilde{\bs \psi}}}

\def\R{\ensuremath{\mathbb{R}}}
\def\N{\ensuremath{\mathcal{N}}}
\def\E{\ensuremath{\mathbb{E}}}

\def\MSD{\text{MSD}}
\def\col{\text{col}}
\def\diag{\text{diag}}
\def\tr{\text{tr}}

\begin{document}

%
\title{Transient Theoretical Analysis of Diffusion RLS Algorithm for Cyclostationary Colored Inputs}
%
%
%

\author{Wei~Gao,~\IEEEmembership{Member,~IEEE},
        Jie~Chen,~\IEEEmembership{Senior Member,~IEEE}, and
        C\'edric~Richard~\IEEEmembership{Senior Member,~IEEE} \\

\thanks{Manuscript received MM DD, 2021; revised MM DD, 2021.}
\thanks{This work was supported in part by the National NSFC under Grant 61701200 and Grant 61671382.}
\thanks{W. Gao is with the School of Computer Science and Telecommunication Engineering, Jiangsu University, Zhenjiang $212013$, China (email: wei\_gao@ujs.edu.cn).}
\thanks{J. Chen is with the Center of Intelligent Acoustics and Immersive Communications (CIAIC), School of Marine Science and Technology, Northwestern Polytechnical University, Xi'an $710072$, China (email: dr.jie.chen@ieee.org).}
\thanks{C. Richard is with the Universit\'e C\^ote d'Azur, OCA, CNRS, 06108 Nice, France (e-mail: cedric.richard@unice.fr).}}

\maketitle

\begin{abstract}
Convergence of the diffusion RLS (DRLS) algorithm to steady-state has been extensively studied in the literature, whereas no analysis of its transient convergence behavior has been reported yet. In this letter, we conduct a theoretical analysis of the transient behavior of the DRLS algorithm for cyclostationary colored inputs, in the mean and mean-square error sense. The resulting analytical models allows us to thoroughly investigate the convergence behavior of the algorithm over adaptive networks in such complex scenarios. Simulation results support the accuracy and correctness of the theoretical findings.
\end{abstract}


%

\section{Introduction}
\label{sec:Intro}

\IEEEPARstart{O}{nline} learning with streaming data in a distributed and collaborative manner can be useful in a wide range of applications~\cite{cooperative2018djuric}. This topic has been receiving considerable attention in recent years with emphasis on both  single-task~\cite{6777576} and multitask diffusion strategies~\cite{nassif2019multitask,chen2014multitask,chen2016multitask}.
Among existing diffusion strategies, extensive studies have been conducted on the derivation of distributed RLS-type algorithms due to their well-appreciated superiority over the non-cooperative RLS.

The low communications distributed RLS (LC-dRLS) algorithm was primitively proposed in~\cite{Sayed2006}, and its steady-state mean-square performance analysis conducted. The diffusion RLS (DRLS) algorithm via incremental update was developed to address the problem of distributed estimation~\cite{Cattivelli2008}. Its steady-state mean-square error was also analyzed in this paper. In~\cite{Bertrand2011}, diffusion adaptation strategy was applied to bias-compensated RLS algorithm for reducing the residual bias. The authors also derived closed-form expressions describing the steady-state mean and mean-square performances. Several distributed sparse RLS algorithms were presented in~\cite{Li2014} along with their steady-state mean and mean-square performance. The distributed sparse multitask RLS  was recently studied in~\cite{Cao2017}. The partial DRLS (PDRLS) algorithm was derived to achieve a trade-off between estimation accuracy and communication burden~\cite{Arablouei2014}. In addition, the convergence performance of the PDRLS was analyzed in both mean and mean-square senses using energy conservation argument. More recently, work~\cite{Yu2019} introduced the robust DRLS algorithm to improve convergence performance in the presence of impulsive noise. The reduced-communication DRLS algorithm and its steady-state analytical models were provided in~\cite{Rastegarnia2020}.

Almost all existing works that report analyses of DRLS-type algorithms using the energy conservation principle only focus on the steady-state mean and mean-square error performances under the assumption of stationary inputs. Nevertheless, cyclostationary inputs with periodical variations are ubiquitous in real-world applications~\cite{Gardner1991, Gardner2006}. In that way, the theoretical performance of LMS-type algorithms were studied further based on this assumption~\cite{Bershad2014, Eweda2017, Eweda2018, Bermudez2019, Bershad2019, Shlezinger2019}. Then, the convergence behavior of the diffusion LMS (DLMS) was also analyzed for cyclostationary inputs in~\cite{Wang2018, Gao2019access, Gao2020}. Inspired by a recent work analyzing the non-cooperative RLS algorithm with new tools~\cite{Eweda2020}, in this letter we propose a theoretical analysis of the transient behavior of the DRLS algorithm over adaptive networks for cyclostationary colored inputs. To the best of our knowledge, this is the first work which provides insight into the transient convergence behavior of the DRLS in such scenarios. More importantly, our work proposes a standard framework that can be readily applied to the transient analysis of other DRLS-type algorithms. Simulation results show the accuracy and correctness of the theoretical findings.


\section{Preliminaries and DRLS algorithm}
\label{sec:Preliminaries}


\subsection{Distributed system model}
\label{subsec:Model}

Consider an adaptive network consisting of $K$ nodes with some communication links between these nodes. Every node $k$ has access to a sequence of the collected input-output data pairs $\{\vx_{k,n}, d_{k,n}\}_{n=1}^N$, where $\vx_{k,n}=[x_{k,n}, \ldots, x_{k,n-L+1}]^\top$ is the input regression vector and $d_{k,n}$ is the scalar desired response assumed to be zero-mean. At time instant $n$ for each node $k$, the desired output $d_{k,n}$ is assumed to be generated from the input vector $\vx_{k,n}$ via the linear regression model
\begin{equation}
    \label{eq:model}
    d_{k,n} = \vx_{k,n}^\top \vw^\star + z_{k,n}
\end{equation}
where $\vw^\star\in\R^L$ is the optimal weight vector to be estimated, and $z_{k,n}$ denotes the zero-mean Gaussian observation noise with variance $\sigma_{z,k}^2$. Specifically, $z_{k,n}$ is assumed to be independent of any other signals. In addition, the cyclostationary input vector $\vx_{k,n}$ at node $k$ and time instant $n$ is modeled as a colored random process with the periodically time-varying variance~\cite{Eweda2020}:
\begin{equation}
    \label{eq:Cyclostationary}
    x_{k,n} = \sigma_{x,k}(n) \, u_{k,n}
\end{equation}
where $\sigma_{x,k}(n)$ is a deterministic periodic sequence with repetition period $T$, and $u_{k,n}$ is a zero-mean colored Gaussian random sequence with variance $\sigma_{u,k}^2=1$. If not the most general one, model~\eqref{eq:Cyclostationary} is a typical form of cyclostationary signal used in many practical applications.
The time-varying autocorrelation matrix of input vector $\vx_{k,n}$ in~\eqref{eq:Cyclostationary} is given by:
\begin{equation}
    \label{eq:mRxk}
    \mR_{x,k}(n) = \E\{\vx_{k,n} \vx_{k,n}^\top\} = \mSigma_{x,k}(n) \mR_{u,k} \mSigma_{x,k}(n)
\end{equation}
with $\mSigma_{x,k}(n)=\diag\big\{\sigma_{x,k}(n), \ldots, \sigma_{x,k}(n-L+1)\big\}$ a diagonal matrix, and $\mR_{u,k}=\E\{\vu_{k,n} \vu_{k,n}^\top\}$ the autocorrelation matrix. The sinusoidal and pulsed variation models~\cite{Bershad2014, Eweda2017, Eweda2018, Bermudez2019, Bershad2019, Shlezinger2019} are often adopted for the periodic sequences $\sigma_{x,k}^2(n)$ to study the impact of cyclostationary colored inputs on transient behavior.

\subsection{Diffusion RLS algorithm}
\label{subsec:DRLS}

Distributed adaptive filtering algorithms are usually implemented using two common diffusion strategies, the adapt-then-combine (ATC) and the combine-then-adapt (CTA)~\cite{Sayed2014Proc, Sayed2014book}. Subsequently, we shall focus on the DRLS with ATC diffusion strategy by sequentially solving two local least-squares problems based on local measurements, which correspond to the adaptation and combination steps, respectively.

Let us first define the local estimate $\vw_{k,n-1}$ and intermediate estimate $\vpsi_{k,n}$ of $\vw^\star$ at node $k$ and time $n$, respectively. Since $\vw_{k,n-1}$ is a good guess for $\vpsi_{k,n}$ as a prior information, we thus solve the following local cost function:
\begin{equation}
    \label{eq:CostFun1}
    \vpsi_{k,n} = \mathop{\arg\min}\limits_{\vpsi\in\R^L} \big\{ \| \vpsi - \vw_{k,n-1} \|^2_{\mLambda_{k,n}} + ( d_{k,n} - \vx_{k,n}^\top \vpsi )^2 \big\}
\end{equation}
with $\mLambda_{k,n} = \mPhi_{k,n} - \vx_{k,n} \vx_{k,n}^\top$ a positive-definite weighting matrix, and $\mPhi_{k,n}$ the time-averaged correlation matrix of input data for node $k$ defined by~\cite{Yu2019}:
\begin{equation}
    \label{eq:mPhi}
    \mPhi_{k,n} = \lambda \mPhi_{k,n-1} + \vx_{k,n} \vx_{k,n}^\top
\end{equation}
with forgetting factor $0\ll\lambda<1$ and initial condition $\mPhi_{k,0}=\delta\mI_L$ with a small positive value $\delta$. By substitution of variables, i.e., $\vv=\vpsi - \vw_{k,n-1}$ and $b=d_{k,n}-\vx_{k,n}^\top \vw_{k,n-1}$, we get:
\begin{equation}
    \label{eq:CostFun2}
    \vv^\star = \mathop{\arg\min}\limits_{\vv\in\R^L} \big\{ \vv^\top \mLambda_{k,n} \vv + ( b - \vx_{k,n}^\top \vv )^2 \big\}.
\end{equation}
Setting the derivative of~\eqref{eq:CostFun2} with respect to $\vv$ to zero, we then arrive at the solution:
\begin{equation}
    \label{eq:Adaptation}
    \vpsi_{k,n} = \vw_{k,n-1} + \mP_{k,n}  \vx_{k,n} e_{k,n}
\end{equation}
with $\mP_{k,n} = \mPhi_{k,n}^{-1}$ and $e_{k,n}$ the instantaneous estimation error at node $k$ and time instant $n$, i.e.,
\begin{equation}
    \label{eq:ekn}
    e_{k,n} = d_{k,n} - \vx_{k,n}^\top \vw_{k,n-1}.
\end{equation}
Applying the matrix inversion lemma to the r.h.s. of~\eqref{eq:mPhi}, the update equation of the inverse of autocorrelation matrix at node $k$ is given by~\cite{Haykin1991, Sayed2003}
\begin{equation}
    \label{eq:mP}
    \mP_{k,n} = \lambda^{-1} \bigg( \mP_{k,n-1} - \frac{\lambda^{-1} \mP_{k,n-1} \vx_{k,n} \vx_{k,n}^\top \mP_{k,n-1}}{1 + \lambda^{-1} \vx_{k,n}^\top \mP_{k,n-1} \vx_{k,n}} \bigg)
\end{equation}
with the initial condition $\mP_{k,0}=\delta^{-1}\mI_L$.

We collect the intermediate estimates $\vpsi_{k,n}$ for all nodes into a column vector of length $KL$, and stack $K$ identity matrices on top of each other into a $KL\times L$ matrix, as follows:
\begin{equation}
    \label{eq:vpsibi}
    \vpsi_n = \col\{ \vpsi_{1,n}, \ldots, \vpsi_{K,n} \}, \quad \bI = \col\{ \mI_L, \ldots, \mI_L \}.
\end{equation}
We then attempt to further improve the estimation precision of $\vpsi_{k,n}$ for each node $k$ by sharing local data available within its neighborhood. Given the entire intermediate estimate $\vpsi_n$, we solve the following weighted least squares problem~\cite{Cattivelli2008}:
\begin{equation}
    \label{eq:LS2}
    \vw_{k,n} = \mathop{\arg\min}\limits_{\vw\in\R^L} \{ \|\vpsi_n - \bI \vw\|_{\mPi_k}^2 \}
\end{equation}
with $\mPi_k = \diag\big\{ a_{1k} \mI_L, \ldots, a_{Kk} \mI_L \big\}$ the node-dependent weighting block diagonal matrix, and $\{a_{\ell k}\}_{k=1}^K$ the set of non-negative combination coefficients~\cite{Sayed2014Proc, Sayed2014book}. Each coefficient $a_{\ell k}$ is the $(\ell, k)$-th entry of a left-stochastic matrix $\mA$, i.e., $\mA^\top\1_K=\1_K$. Likewise, setting the derivative of~\eqref{eq:LS2} with respect to $\vw$ to zero, we finally find the solution
\begin{equation}
    \label{eq:Combination}
    \vw_{k,n} = \sum_{\ell\in \N_k} a_{\ell k} \, \vpsi_{\ell,n}.
\end{equation}
Even though the local intermediate estimates $\vpsi_{k,n}$ may suffer from random estimation error fluctuations, the combined solution~\eqref{eq:Combination} is able to achieve better estimation accuracy. Note that the weight update equation~\eqref{eq:Adaptation} and recursion~\eqref{eq:mP} with combination~\eqref{eq:Combination} still offer good estimation performance instead of implementing an incremental-type scheme as in~\cite{Cattivelli2008}. This will be illustrated by simulation results later.

\section{Transient Performance Analysis}
\label{sec:Transient}

We now study the transient performance analysis of DRLS with ATC diffusion strategy for cyclostationary colored inputs in the mean and mean-square sense. The error vectors for node $k$ at time instant $n$ are defined respectively as:
\begin{equation}
    \label{eq:error0}
    \hatvpsi_{k,n} \triangleq \vpsi_{k,n} - \vw^\star, \quad \hatvw_{k,n} \triangleq \vw_{k,n} - \vw^\star.
\end{equation}
Then, let $\hatvpsi_n$ and $\hatvw_n$ denote the block weight error vectors by collecting the error vectors for all nodes as follows:
\begin{equation}
    \label{eq:error1}
    \hatvpsi_n \triangleq \col \{\hatvpsi_{1,n}, \ldots, \hatvpsi_{K,n} \}, \quad \hatvw_n \triangleq \col \{\hatvw_{1,n}, \ldots, \hatvw_{K,n} \}.
\end{equation}
We introduce the following $K\times K$ block diagonal matrices with individual entries of size $L\times L$: 
\begin{align}
    \label{eq:mRn}
    \bR_{x,n} &\triangleq \diag\{ \vx_{1,n} \vx_{1,n}^\top, \ldots, \vx_{K,n} \vx_{K,n}^\top \} \\
    \label{eq:mPn}
    \mP_n &\triangleq \diag\{\mP_{1,n}, \ldots, \mP_{K,n} \} \\
    \label{eq:mPhin}
    \mPhi_n &\triangleq \diag\{\mPhi_{1,n}, \ldots, \mPhi_{K,n} \}
\end{align}
and the $K\times1$ block column vector with vectors of length $L$:
\begin{equation}
    \label{eq:mMn}
    \vs_{xz,n} \triangleq \col\{z_{1,n} \vx_{1,n}, \ldots, z_{K,n} \vx_{K,n} \}.
\end{equation}

Before proceeding, we introduce the following independence assumption.

\textbf{A1.} The regression input random vector $\vx_{k,n}$ is modeled by a cyclostationary random process that is temporally and spatially independent with covariance matrix $\mR_{x,k}(n)$.

\subsection{Mean weight error analysis}
\label{subsec:MeanAnalysis}

Collecting both sides of~\eqref{eq:mPhi} for each nodes, using~\eqref{eq:mRn} and~\eqref{eq:mPhin}, we obtain:
\begin{equation}
    \label{eq:mPhi0}
    \mPhi_n  = \lambda \mPhi_{n-1} + \bR_{x,n}.
\end{equation}
Then taking expectations of both sides leads to:
\begin{equation}
    \label{eq:EmPhi}
    \E\{ \mPhi_n \} = \lambda \E\{ \mPhi_{n-1} \} + \mR_{x,n}
\end{equation}
with the initial condition $\E\{ \mPhi_0 \}=\delta^{-1}\diag\{\mI_L, \ldots, \mI_L \}$ and $\mR_{x,n}=\E\{\bR_{x,n}\}=\diag\big\{\mR_{x,1}(n), \ldots, \mR_{x,K}(n)\big\}$ a block diagonal matrix. Note that $\mPhi_n$ only depends on $\mR_{x,n}$. The recursive relation~\eqref{eq:EmPhi} is very useful and crucial in the following theoretical analysis. 

Replacing~\eqref{eq:model} into~\eqref{eq:ekn} and using definition $\hatvw_{k,n}$, the instantaneous estimation error can be rewritten as:
\begin{equation}
    \label{eq:error5}
    e_{k,n} = z_{k,n} - \vx_{k,n}^\top \hatvw_{k,n-1}.
\end{equation}
Subtracting $\vw^\star$ from both sides of the recursive update equation~\eqref{eq:Adaptation}, then using~\eqref{eq:error0} and~\eqref{eq:error5}, leads to the intermediate weight error vector update equation:
\begin{equation}
    \label{eq:hatvpsi_k}
    \hatvpsi_{k,n} = \hatvw_{k,n-1} + \mP_{k,n} \vx_{k,n} ( z_{k,n} - \vx_{k,n}^\top \hatvw_{k,n-1} ).
\end{equation}
Subtracting $\vw^\star$ from both sides of combination relation~\eqref{eq:Combination}, and using~\eqref{eq:error0}, we have the weight error vector:
\begin{equation}
    \label{eq:hatvw_k}
    \hatvw_{k,n} = \sum_{\ell\in \N_k} a_{\ell k} \, \hatvpsi_{\ell,n}.
\end{equation}
Substituting~\eqref{eq:hatvpsi_k} into~\eqref{eq:hatvw_k}, and using the previously introduced expressions~\eqref{eq:error1}--\eqref{eq:mPn} and~\eqref{eq:mMn}, then the recursive update equation of global weight error vector can be formulated as:
\begin{equation}
    \label{eq:UpdateEq1}
        \hatvw_n = \bA\big( \hatvw_{n-1} - \mP_n \bR_{x,n} \hatvw_{n-1} + \mP_n \vs_{xz,n} \big)
\end{equation}
with the matrix $\bA=\mA^\top\otimes\mI_L$. Pre-multiplying both sides of~\eqref{eq:UpdateEq1} by $\mP^{-1}_n\bA^{-1}$, and using~\eqref{eq:mPhi0}, it follows that:
\begin{equation}
    \label{eq:UpdateEq2}
    \mPhi_n \bA^{-1} \hatvw_n = \lambda \mPhi_{n-1} \hatvw_{n-1} + \vs_{xz,n}.
\end{equation}
The purpose of these matrix manipulations was to remove matrix $\mP_n$ that multiplies vector $\vs_{xz,n}$ on the r.h.s. of~\eqref{eq:UpdateEq1}, which results in the mathematical intractability of the mean weight error analysis. Taking expectation of both sides of~\eqref{eq:UpdateEq2} and applying the statistical properties of noise $z_{k,n}$, we obtain:
\begin{equation}
    \label{eq:UpdateEq3}
    \E\{ \mPhi_n \bA^{-1} \hatvw_n \} = \lambda \E\{\mPhi_{n-1} \hatvw_{n-1} \}.
\end{equation}
For the sake of mathematical tractability, we follow~\cite{Eweda2020} where the authors show that the following two approximations hold for cyclostationary inputs:
\begin{align}
    \label{eq:ApproxPM1}
    \E\{ \mPhi_n \bA^{-1} \hatvw_n \} &\approx \E\{ \mPhi_n \} \bA^{-1} \E\{ \hatvw_n \} \\
    \label{eq:ApproxPM2}
    \E\{ \mPhi_n \hatvw_n \} &\approx \E\{ \mPhi_n \} \E\{ \hatvw_n \}.
\end{align}
To understand these two approximations, we further assume that $\mDelta_n$ is the random fluctuation of $\mPhi_n$ around $\E\big\{\mPhi_n\big\}$, which can be expressed as~\cite{Eweda2020}:
\begin{equation}
    \label{eq:mDelta}
    \mPhi_n = \E\big\{ \mPhi_n \big\} + \mDelta_n.
\end{equation}
Correspondingly,~\eqref{eq:ApproxPM1} and~\eqref{eq:ApproxPM2} can be written as:
\begin{equation}
    \label{eq:ApproxPM3}
    \E\big\{ \mPhi_n \bA^{-1} \hatvw_n \big\} \!=\! \E\big\{ \! \mDelta_n \bA^{-1} \hatvw_n \! \big\} \!+\! \E\big\{ \! \mPhi_n \! \big\} \bA^{-1} \E\big\{ \! \hatvw_n \! \big\}
\end{equation}
\begin{equation}
    \label{eq:ApproxPM4}
    \E\big\{ \mPhi_n \hatvw_n \big\} = \E\big\{ \mDelta_n \hatvw_n \big\} + \E\big\{ \mPhi_n \big\} \E\big\{ \hatvw_n \big\}.
\end{equation}
As shown in~\cite{Eweda2020}, since random matrix $\mDelta_n$ is small with respect to $\E\big\{\mPhi_n\big\}$ for cyclostationary inputs, the first terms on the r.h.s. of~\eqref{eq:ApproxPM3} and~\eqref{eq:ApproxPM4} are small enough to be eliminated~\cite{Eweda2020}, respectively. Substituting the approximations~\eqref{eq:ApproxPM1} and~\eqref{eq:ApproxPM2} into~\eqref{eq:UpdateEq3}, yields:
\begin{equation}
    \label{eq:UpdateEq4}
    \E\{\mPhi_n\} \bA^{-1}\E\{\hatvw_n\} = \lambda \, \E\{ \mPhi_{n-1}\} \E\{\hatvw_{n-1}\}.
\end{equation}
Pre-multiplying both sides of~\eqref{eq:UpdateEq4} by $\bA\,\E\{\mPhi_n\}^{-1}$ leads to the mean weight error behavior of DRLS algorithm, given by:
\begin{equation}
    \label{eq:UpdateEq5}
    \E\{ \hatvw_n \} = \lambda \, \bA\, \E\{\mPhi_n\}^{-1} \E\{ \mPhi_{n-1}\} \E\{ \hatvw_{n-1} \}.
\end{equation}

\noindent\textit{Theorem 1} (Convergence in the mean) Given a left-stochastic matrix $\mA$, the weight error vector of the DRLS algorithm with ATC strategy converges to a zero vector as $n\to\infty$, that is,
\begin{equation}
    \label{eq:MeanStability}
    \vspace{-1mm}
    \lim_{n\to\infty} \E\{ \hatvw_n \} = \0_{KL}
\end{equation}
which ensures that the estimate of DRLS algorithm is asymptotically unbiased and converge in the mean sense, i.e., $\lim_{n\to\infty}\E\{\vw_{k,n}\}=\vw^\star$ for all nodes $k$.

\textit{Proof:} Since $\E\{\mPhi_n\}^{-1} \E\{ \mPhi_{n-1}\} \approx \mI_{KL}$ as $n\to\infty$ and $0\ll\lambda <1$, we can conclude from~\eqref{eq:UpdateEq5}. \hfill $\blacksquare$

\vspace{-1mm}
\subsection{Mean-square weight error analysis}
\label{subsec:MeanSquareAnalysis}

We shall now proceed with the mean-square transient analysis of the DRLS algorithm by considering the network mean-square deviation (MSD) defined as $\MSD_n = \tr\big\{ \mK_n \big\}/K$, where $\mK_n=\E\big\{\hatvw_n \hatvw^\top_n\big\}$ is the second-order moment matrix of the global weight error vector across all nodes~\cite{Sayed2014book, Sayed2014Proc}. Hence, our aim is to determine the recursive update equation of $\mK_n$ to evaluate the network transient MSD.

Post-multiplying~\eqref{eq:UpdateEq2} by its transpose, and taking the expectation of its both sides, we obtain:
\begin{equation}
    \label{eq:Km1}
    \begin{split}
    \E\big\{ \mPhi_n \bA^{-1} \hatvw_n \hatvw^\top_n &(\bA^{-1})^\top \mPhi_n \big\} = \E\big\{ \vs_{xz,n} \vs_{xz,n}^\top \big\} \\
    & + \lambda^2 \, \E\big\{ \mPhi_{n-1} \hatvw_{n-1} \hatvw^\top_{n-1} \mPhi_{n-1} \big\}
    \end{split}
\end{equation}
where the zero-valued cross terms have been eliminated. To make the analysis tractable, we introduce two approximations that hold for cyclostationary inputs~\cite{Eweda2020}:
\begin{equation}
    \label{eq:ApproxSec1}
    \begin{split}
    &\E\big\{ \mPhi_n \bA^{-1} \hatvw_n \hatvw^\top_n (\bA^{-1})^\top \mPhi_n \big\} \\
    &\qquad\approx \E\{ \mPhi_n \} \bA^{-1} \mK_n (\bA^{-1})^\top \E\{ \mPhi_n \}
    \end{split}
\end{equation}
\begin{equation}
    \label{eq:ApproxSec2}
    \E\big\{ \mPhi_n \hatvw_n \hatvw^\top_n \mPhi_n \big\} \approx \E\big\{ \mPhi_n \big\} \mK_n \E\big\{ \mPhi_n \big\}.
\end{equation}
To construct these two approximations, we use~\eqref{eq:mDelta} with the l.h.s. terms of \eqref{eq:ApproxSec1} and~\eqref{eq:ApproxSec2}:
\begin{align}
    \label{eq:ApproxSec4}
    &\E\big\{ \mPhi_n \bA^{-1} \hatvw_n \hatvw^\top_n (\bA^{-1})^\top \mPhi_n \big\} = \nonumber\\
    &\quad \E\{ \mPhi_n \} \bA^{-1} \mK_n (\bA^{-1})^\top\E\{ \mPhi_n \} \nonumber \\
    &\quad + \E\{ \mPhi_n \} \E\big\{\bA^{-1} \hatvw_n \hatvw^\top_n (\bA^{-1})^\top \mDelta_n \big\} \nonumber \\
    &\quad+ \E\big\{ \mDelta_n \bA^{-1} \hatvw_n \hatvw^\top_n (\bA^{-1})^\top \big\} \E\{ \mPhi_n \} \\
    &\quad+ \E\big\{ \mDelta_n \bA^{-1} \hatvw_n \hatvw^\top_n (\bA^{-1})^\top \mDelta_n \big\} \nonumber \\
    \label{eq:ApproxSec5}
    &\E\big\{ \mPhi_n \hatvw_n\hatvw^\top_n \mPhi_n \big\} = \E\big\{ \mPhi_n \big\} \mK_n \E\big\{ \mPhi_n \big\} \nonumber \\
    & + \E\big\{ \mDelta_n \hatvw_n \hatvw^\top_n \mDelta_n \big\} \nonumber \\
    & + \E\big\{ \mDelta_n \hatvw_n \hatvw^\top_n \big\} \E\big\{ \mPhi_n \big\}
    + \E\big\{ \mPhi_n \big\} \E\big\{ \hatvw_n \hatvw^\top_n \mDelta_n \big\}.
\end{align}
Similarly to the justifications for approximations~\eqref{eq:ApproxPM1} and~\eqref{eq:ApproxPM2}, observe that~\eqref{eq:ApproxSec4} and~\eqref{eq:ApproxSec5} reasonably give rise to approximations~\eqref{eq:ApproxSec1} and~\eqref{eq:ApproxSec2}. See~\cite{Eweda2020} for more details. Their rationality and effectiveness will be validated by simulation results later. Substituting approximations~\eqref{eq:ApproxSec1} and~\eqref{eq:ApproxSec2} into~\eqref{eq:Km1}, yields:
\begin{equation}
    \label{eq:Km2}
    \begin{split}
    &\E\{ \mPhi_n \} \bA^{-1} \mK_n (\bA^{-1})^\top \E\{ \mPhi_n \} \\
    &= \lambda^2 \, \E\{ \mPhi_{n-1} \} \mK_{n-1} \E\{ \mPhi_{n-1} \} + \mS_{xz,n}.
    \end{split}
\end{equation}
Considering assumption A1 and the statistical independence of observation noise $z_{k,n}$, we obtain:
\begin{align}
    \label{eq:mT2}
    \mS_{xz,n} &= \E\big\{ \diag \{z_{1,n}^2 \vx_{1,n} \vx_{1,n}^\top, \ldots, z_{K,n}^2 \vx_{K,n} \vx^\top_{K,n} \} \big\} \\
    &= \diag \big\{\sigma_{z,1}^2 \mR_{x,1}(n), \ldots, \sigma_{z,K}^2 \mR_{x,K}(n) \big\} = \mSigma_z \mR_{x,n} \nonumber
\end{align}
with the observation noise variance block diagonal matrix $\mSigma_z = \diag \{ \sigma_{z,1}^2 \mI_L, \ldots, \sigma_{z,K}^2 \mI_L \}$.
Pre-multiplying~\eqref{eq:Km2} by $\bA\, \E\{ \mPhi_n \}^{-1}$ and post-multiplying~\eqref{eq:Km2} by $\E\{ \mPhi_n \}^{-1}\bA^\top$ simultaneously, we finally obtain:
\begin{align}
    \label{eq:Km3}
    \mK_n &= \bA \big( \lambda^2 \, \E\{ \mPhi_n \}^{-1} \E\{ \mPhi_{n-1} \} \mK_{n-1} \E\{ \mPhi_{n-1} \} \E\{ \mPhi_n \}^{-1} \nonumber \\
    &\quad + \E\{ \mPhi_n \}^{-1} \mSigma_z \mR_{x,n} \E\{ \mPhi_n \}^{-1} \big)\bA^\top
\end{align}
where the recursive relation~\eqref{eq:EmPhi} is invoked for the update of matrix $\mK_n$. As a consequence,~\eqref{eq:Km3} enables us to investigate the network transient convergence behavior of the DRLS with ATC diffusion strategy in the mean-square sense. In contrast, almost all existing analyses of DRLS-type algorithms are limited to the network steady-state MSD for stationary inputs. More importantly, this work proposes a generic enough framework that can be readily applied to the transient analysis of other DRLS-type algorithms in different scenarios.

\section{Simulation Results}
\label{sec:Experiment}

\setcounter{figure}{1}
\begin{figure*}[!htbp]
	\centering
    \subfigure[Pulsed slow variation ($T=512$).]
	{\includegraphics[trim = 5mm 1mm 9mm 6.5mm, clip, width=0.33\textwidth]{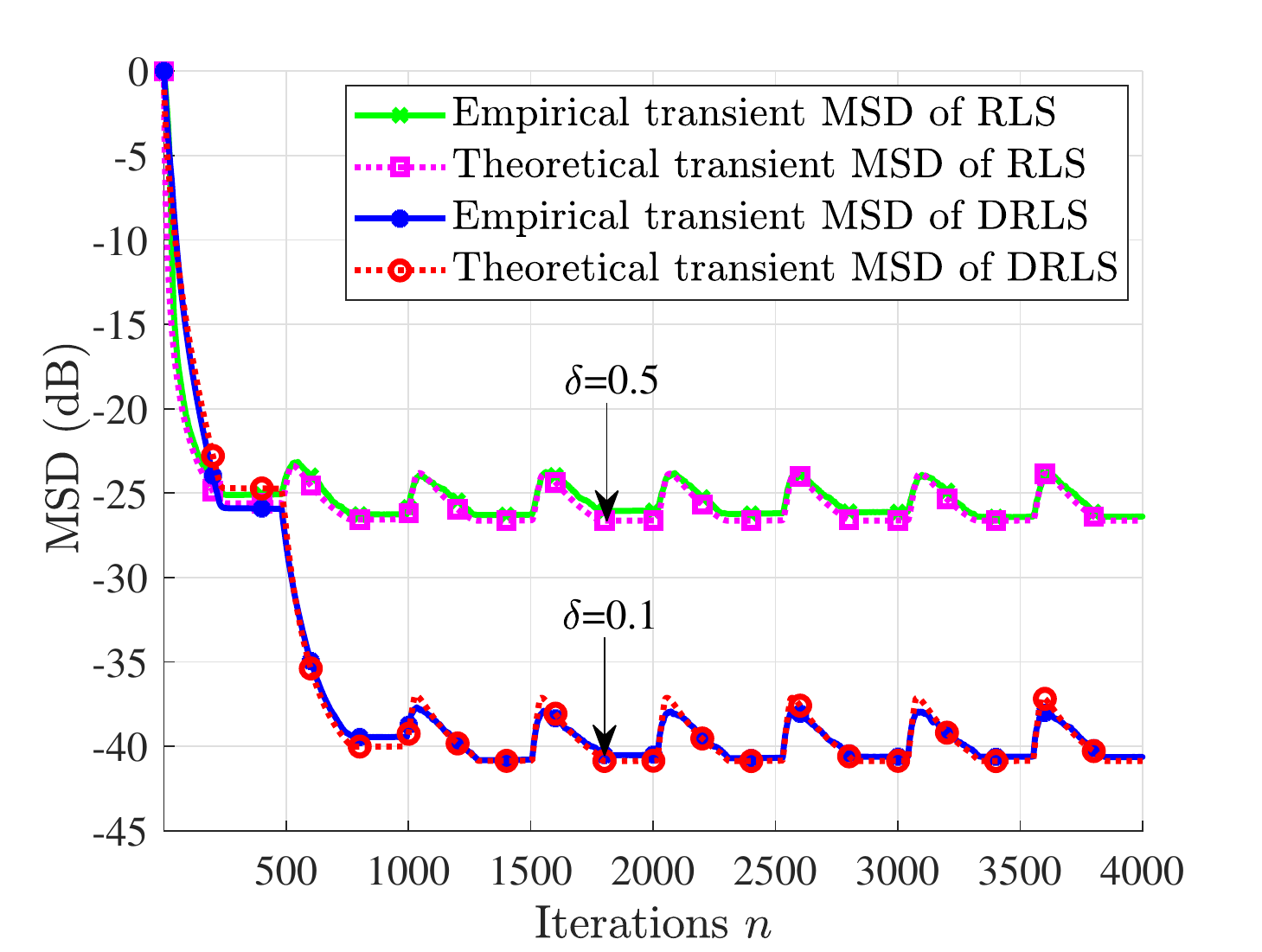}}
    \subfigure[Pulsed moderate variation ($T=32$).]
	{\includegraphics[trim = 5mm 1mm 9mm 6.5mm, clip, width=0.33\textwidth]{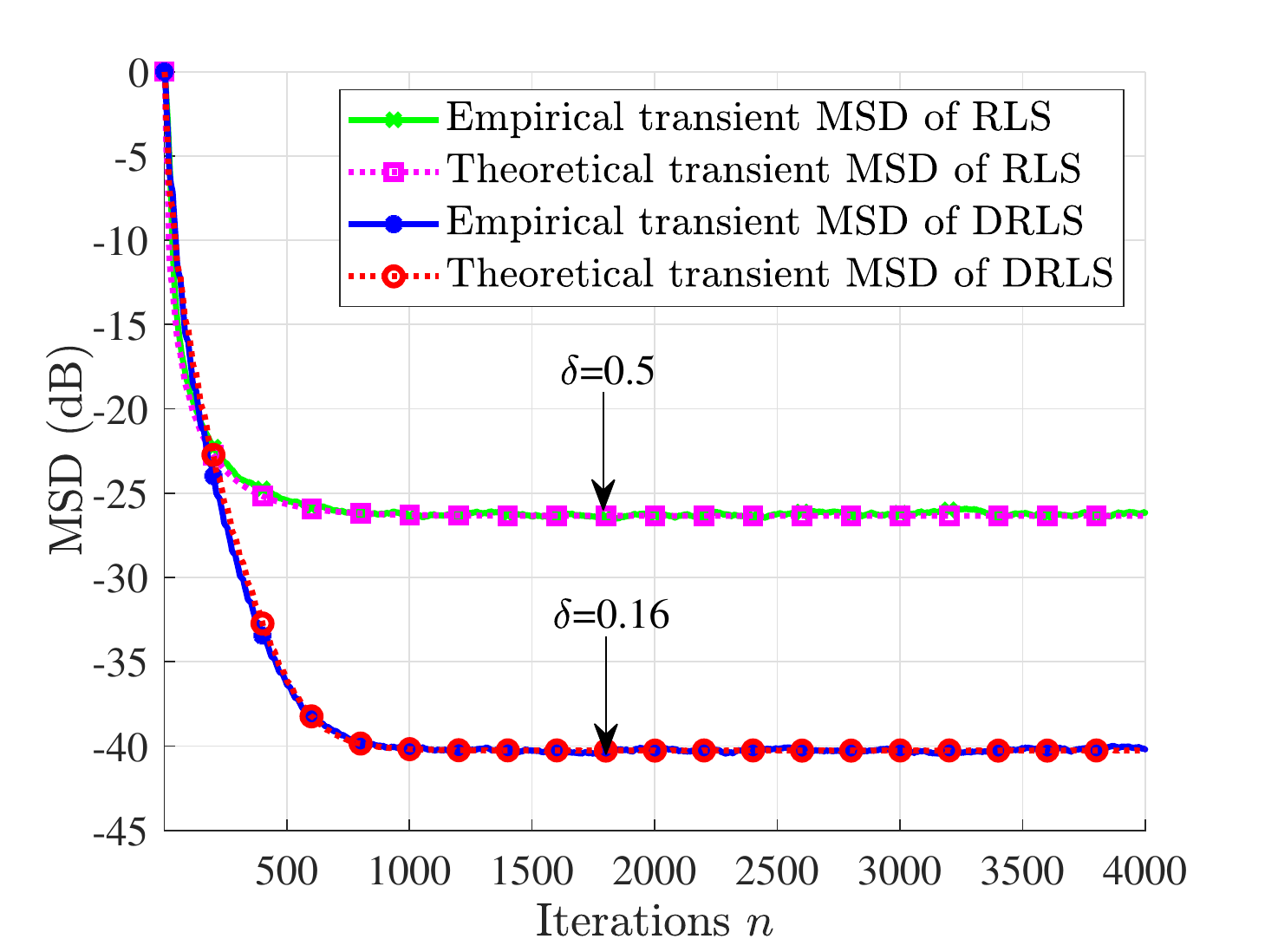}}
    \subfigure[Pulsed fast variation ($T=4$).]
	{\includegraphics[trim = 5mm 1mm 9mm 6.5mm, clip, width=0.328\textwidth]{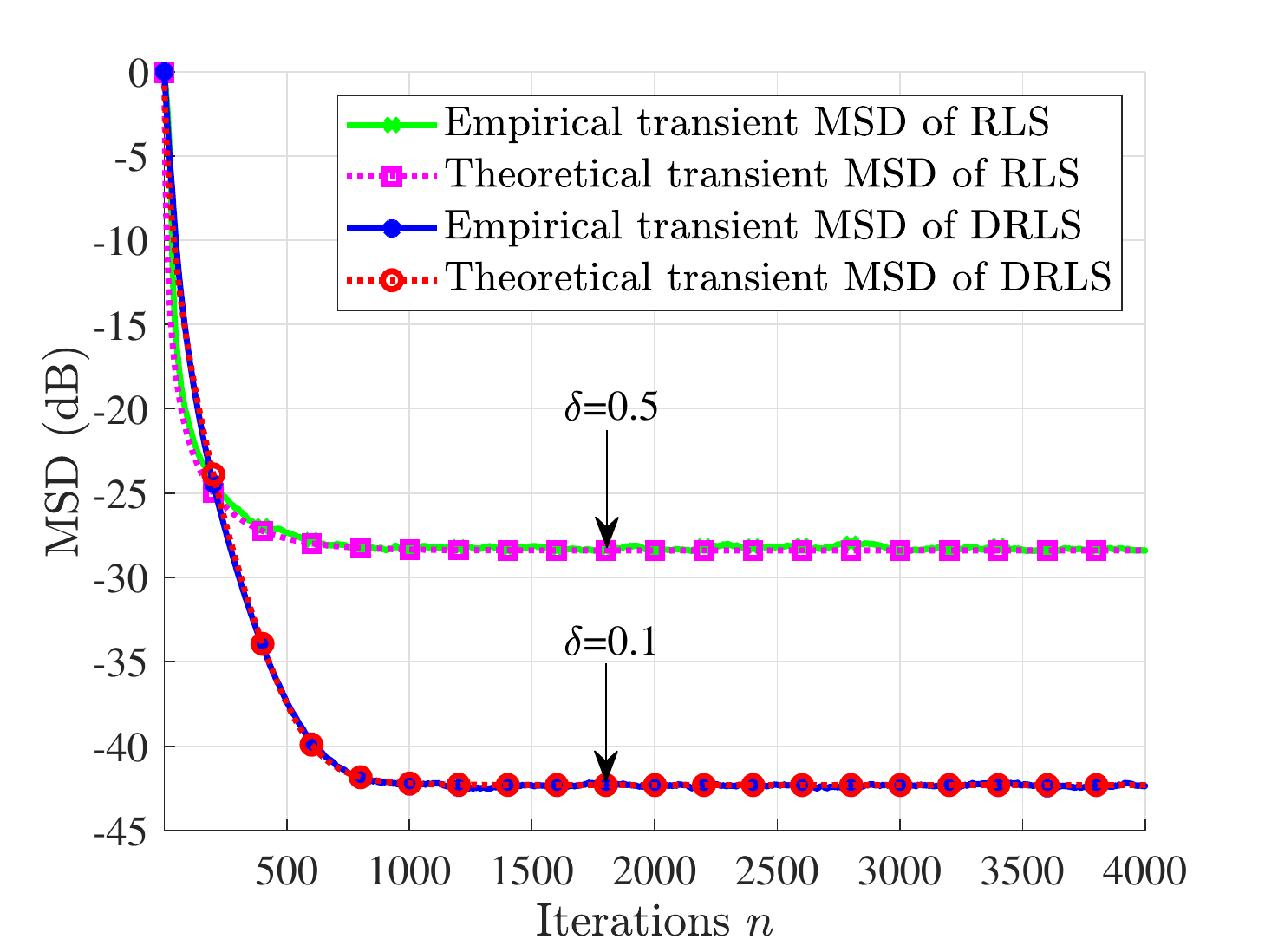}}
    \caption{Comparisons of empirical and theoretical transient MSD of RLS and DRLS algorithms for $\rho=0.8$ and pulsed variations of input variance.}
	\label{fig:Results}
\vspace{-3mm}
\end{figure*}

Consider the network consisting of $20$ nodes with topology illustrated in Fig.~\ref{fig:setup}(a). Figure~\ref{fig:setup}(b) shows the noise variance $\sigma_{z,k}^2$ at each node. The optimal weight vector $\vw^\star \in\R^{32}$ arises from a standard normal distribution scaled by an exponential decaying factor $0.5$, and normalized so that $\|\vw^\star\|^2_2=1$.

\setcounter{figure}{0}
\begin{figure}[!htb]
	\centering
    \subfigure[Network topology.]
	{\raisebox{0.2\height}{\includegraphics[trim = 30mm 29.5mm 23.5mm 33mm, clip, width=0.23\textwidth]{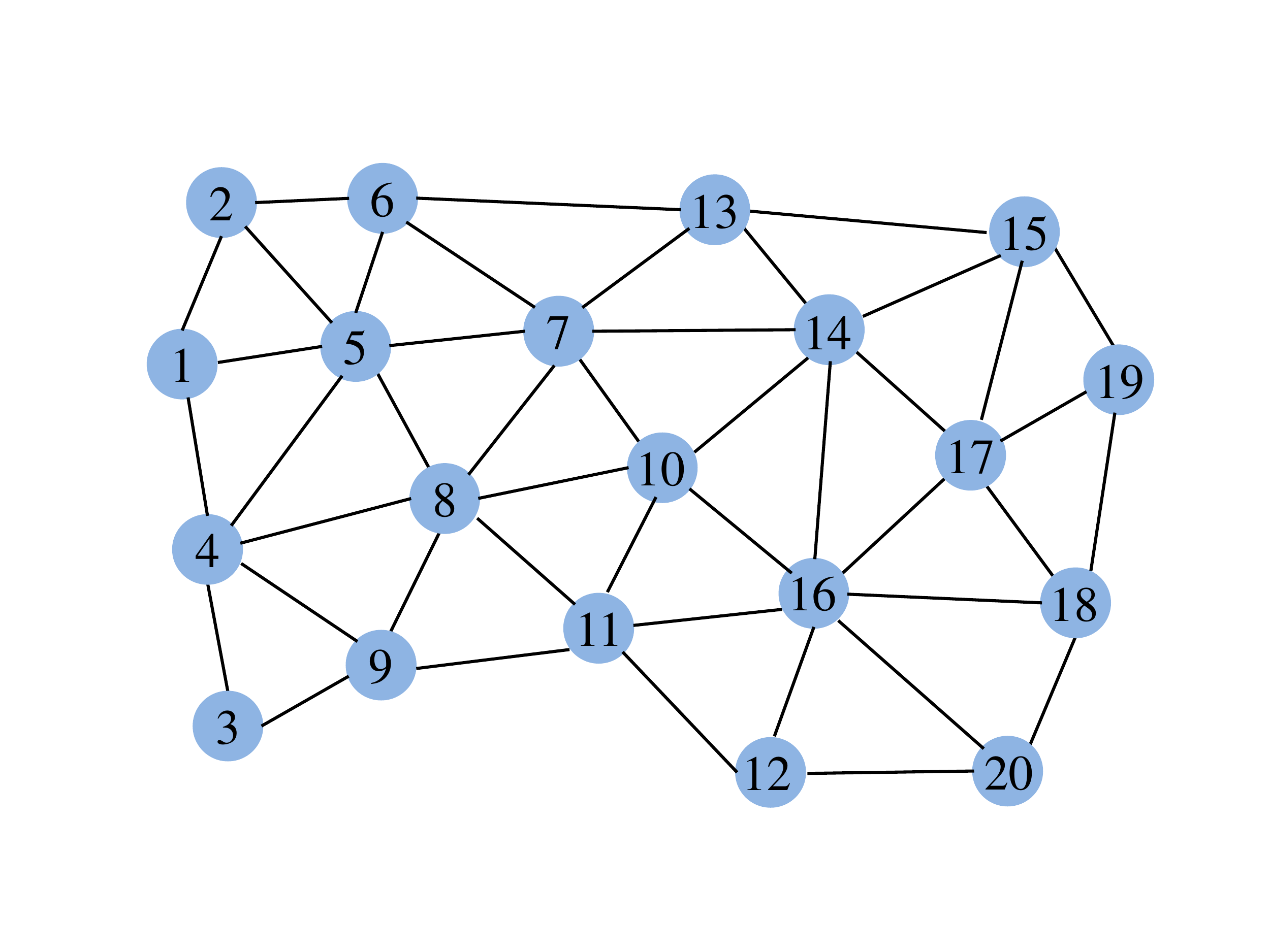}}}
    \subfigure[Noise variances $\sigma_{z,k}^2$.]
	{\includegraphics[trim = 0mm 2mm 11mm 6mm, clip, width=0.25\textwidth]{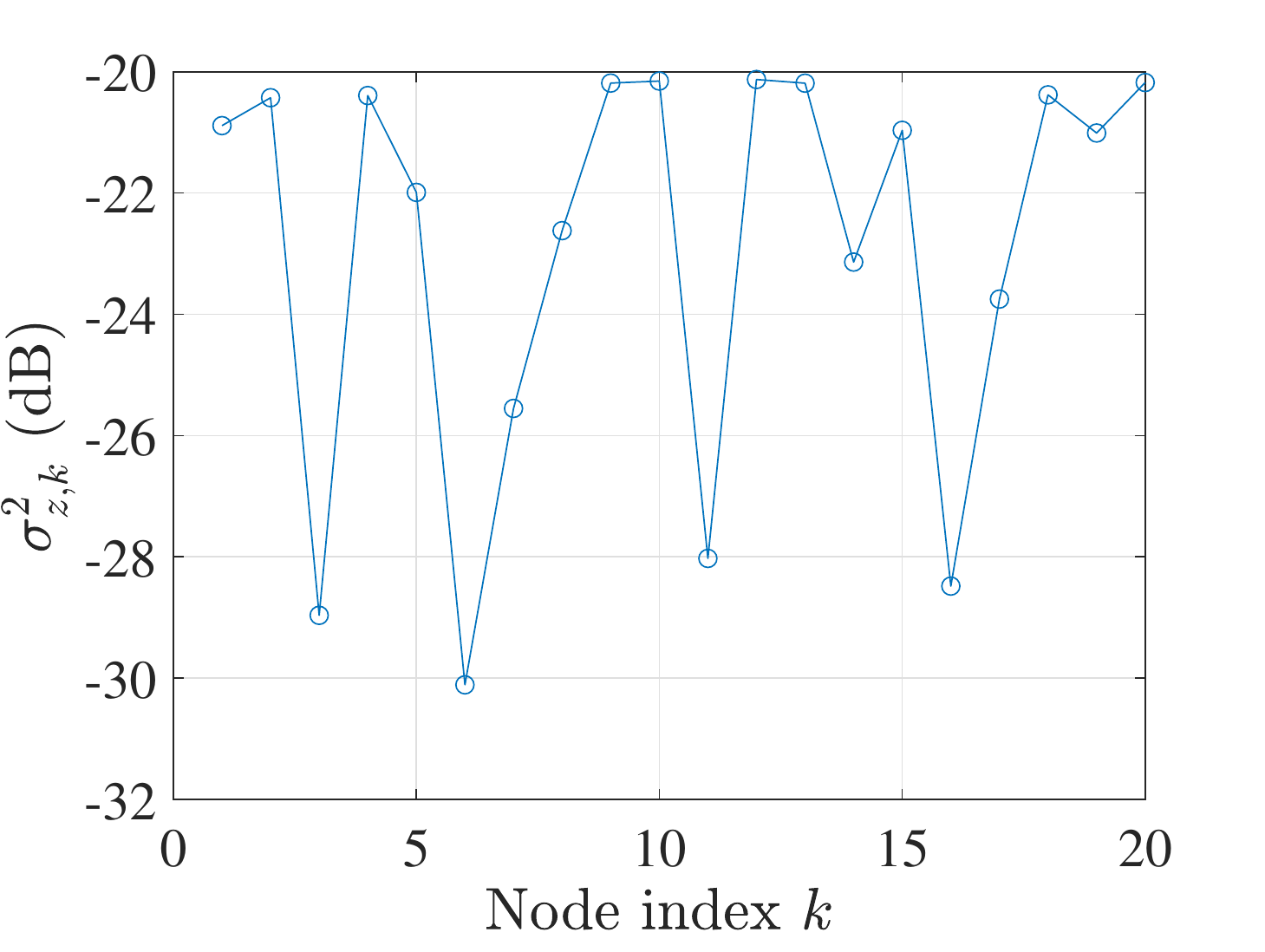}}
    \caption{Experimental setup.}
	\label{fig:setup}
\vspace{-3mm}
\end{figure}

We considered the cyclostationary input model in~\eqref{eq:Cyclostationary}. Each colored Gaussian sequence $u_{k,n}$ was generated independently across all nodes $k$ according to the autoregressive model:
\begin{equation}
    \label{eq:ARmodel}
    u_{k,n} = \rho u_{k,n-1} + \sigma_{u,k} \sqrt{1-\rho^2} w_{k,n}
\end{equation}
with $\rho=0.8$ the normalized correlation factor, and $w_{k,n}$ a zero-mean white Gaussian noise with unit variance. In this way, the $(i,j)$-th entry of autocorrelation matrix of $\vu_{k,n}$ was given by $[\mR_{u,k}]_{ij} = \E\big\{u_{k,n-i+1} u_{k,n-j+1}\big\} = \sigma_{u,k}^2 \rho^{|i-j|}$. A pulsed power time variation model was used for $\sigma_{x,k}(n)$; see~\cite{Gao2020}. The duty cycle $\alpha$, the lowest amplitude $V_l$, and the highest amplitude $V_h$ were set to $0.5$, $2\times10^{-3}$, and $2$, respectively. Three different periods of cyclostationary variations were considered: $T=4$, $32$ and $512$. We set the forgetting factor $\lambda$ to $0.995$ for RLS and DRLS algorithms. Initialization parameter $\delta$ used in~\eqref{eq:EmPhi} was set to different values depending on the scenario; see Fig.~\ref{fig:Results}.
Figure~\ref{fig:Results} shows that the DRLS algorithm with ATC diffusion outperformed the non-cooperative RLS algorithm. We observe that the transient and steady-state MSD periodically fluctuated with the same slow repetition period ($T=512$) as the pulsed input data. On the contrary, the convergence of the DRLS was not affected by the moderate ($T=32$) and fast ($T=4$) variations in the variance of input data. It can be also seen in Fig.~\ref{fig:Results} that the theoretical transient MSD curves of DRLS algorithm predicted by~\eqref{eq:Km3} coincide with the empirical MSD curves, which demonstrates the effectiveness of approximations~\eqref{eq:ApproxSec1} and~\eqref{eq:ApproxSec2}. As a conclusion, the perfect match between theoretical predictions and simulated results demonstrate the accuracy of the analytical models derived in this paper.

\section{Conclusion}
\label{sec:Conclusion}

In this letter, we theoretically analyzed the transient behavior of the DRLS algorithm in the mean and mean-square sense. The derived analytical models allow to thoroughly investigate the variations in the transient convergence behavior. Simulation results illustrated the accuracy and effectiveness of the theoretical findings.



\newpage


\balance

\bibliographystyle{IEEEtran}
\bibliography{DRLS}

%
%

%

%
%
%




\end{document}